\documentstyle[preprint,aps,epsfig]{revtex}

\def \MET {\relax\ifmmode{\mbox{$\raisebox{.3ex}{$\not$}E_T$}}
  \else{$\mbox{$\raisebox{.3ex}{$\not$}E_T$}$~}\fi}
\def \stop {\relax\ifmmode{\tilde{t}_1}\else{$\tilde{t}_1$~}\fi}
\def \stopb {\relax\ifmmode{\bar{\tilde{t}}_1}\else{$\bar{\tilde{t}}_1$~}\fi}
\def \mgev{GeV/c$^2$}
\def \isajet   {{\footnotesize ISAJET}}
\def \vecbos   {{\footnotesize VECBOS}}
\def \herwig   {{\footnotesize HERWIG}}

\begin{document}

\tightenlines
%\draft

\begin{center}
{\large\bf\vskip2.5pt Search for Scalar Top Quark Production in $p\overline{p}$ 
Collisions at $\sqrt{s}=$1.8 TeV}
\end{center}

\font\eightit=cmti8
\def\r#1{\ignorespaces $^{#1}$}
\hfilneg
\begin{sloppypar}
\noindent
T.~Affolder,\r {21} H.~Akimoto,\r {43}
A.~Akopian,\r {36} M.~G.~Albrow,\r {10} P.~Amaral,\r 7 S.~R.~Amendolia,\r {32} 
D.~Amidei,\r {24} K.~Anikeev,\r {22} J.~Antos,\r 1 
G.~Apollinari,\r {36}
% ...
\end{sloppypar}
\vskip .026in
\begin{center}
(CDF Collaboration)
\end{center}

\vskip .026in
\begin{center}
\r 1  {\eightit Institute of Physics, Academia Sinica, Taipei, Taiwan 11529, 
Republic of China} \\
\r 2  {\eightit Argonne National Laboratory, Argonne, Illinois 60439} \\
\r 3  {\eightit Istituto Nazionale di Fisica Nucleare, University of Bologna,
I-40127 Bologna, Italy} \\
\r 4  {\eightit Brandeis University, Waltham, Massachusetts 02254} \\
\r 5  {\eightit University of California at Los Angeles, Los 
Angeles, California  90024} \\  
\r 6  {\eightit Instituto de Fisica de Cantabria, University of Cantabria, 
39005 Santander, Spain} \\
\r 7  {\eightit Enrico Fermi Institute, University of Chicago, Chicago, 
Illinois 60637} \\
\r 8  {\eightit Joint Institute for Nuclear Research, RU-141980 Dubna, Russia}
\\
\r 9  {\eightit Duke University, Durham, North Carolina  27708} \\
\r {10}  {\eightit Fermi National Accelerator Laboratory, Batavia, Illinois 
60510} \\
\r {11} {\eightit University of Florida, Gainesville, Florida  32611} \\
\r {12} {\eightit Laboratori Nazionali di Frascati, Istituto Nazionale di Fisica
               Nucleare, I-00044 Frascati, Italy} \\
\r {13} {\eightit University of Geneva, CH-1211 Geneva 4, Switzerland} \\
\r {14} {\eightit Harvard University, Cambridge, Massachusetts 02138} \\
\r {15} {\eightit Hiroshima University, Higashi-Hiroshima 724, Japan} \\
\r {16} {\eightit University of Illinois, Urbana, Illinois 61801} \\
\r {17} {\eightit The Johns Hopkins University, Baltimore, Maryland 21218} \\
\r {18} {\eightit Institut f\"{u}r Experimentelle Kernphysik, 
Universit\"{a}t Karlsruhe, 76128 Karlsruhe, Germany} \\
\r {19} {\eightit Korean Hadron Collider Laboratory: Kyungpook National
University, Taegu 702-701; Seoul National University, Seoul 151-742; and
SungKyunKwan University, Suwon 440-746; Korea} \\
\r {20} {\eightit High Energy Accelerator Research Organization (KEK), Tsukuba, 
Ibaraki 305, Japan} \\
\r {21} {\eightit Ernest Orlando Lawrence Berkeley National Laboratory, 
Berkeley, California 94720} \\
\r {22} {\eightit Massachusetts Institute of Technology, Cambridge,
Massachusetts  02139} \\   
\r {23} {\eightit Institute of Particle Physics: McGill University, Montreal 
H3A 2T8; and University of Toronto, Toronto M5S 1A7; Canada} \\
\r {24} {\eightit University of Michigan, Ann Arbor, Michigan 48109} \\
\r {25} {\eightit Michigan State University, East Lansing, Michigan  48824} \\
\r {26} {\eightit University of New Mexico, Albuquerque, New Mexico 87131} \\
\r {27} {\eightit The Ohio State University, Columbus, Ohio  43210} \\
\r {28} {\eightit Osaka City University, Osaka 588, Japan} \\
\r {29} {\eightit University of Oxford, Oxford OX1 3RH, United Kingdom} \\
\r {30} {\eightit Universita di Padova, Istituto Nazionale di Fisica 
          Nucleare, Sezione di Padova, I-35131 Padova, Italy} \\
\r {31} {\eightit University of Pennsylvania, Philadelphia, 
        Pennsylvania 19104} \\   
\r {32} {\eightit Istituto Nazionale di Fisica Nucleare, University and Scuola
               Normale Superiore of Pisa, I-56100 Pisa, Italy} \\
\r {33} {\eightit University of Pittsburgh, Pittsburgh, Pennsylvania 15260} \\
\r {34} {\eightit Purdue University, West Lafayette, Indiana 47907} \\
\r {35} {\eightit University of Rochester, Rochester, New York 14627} \\
\r {36} {\eightit Rockefeller University, New York, New York 10021} \\
\r {37} {\eightit Rutgers University, Piscataway, New Jersey 08855} \\
\r {38} {\eightit Texas A\&M University, College Station, Texas 77843} \\
\r {39} {\eightit Texas Tech University, Lubbock, Texas 79409} \\
\r {40} {\eightit Istituto Nazionale di Fisica Nucleare, University of Trieste/
Udine, Italy} \\
\r {41} {\eightit University of Tsukuba, Tsukuba, Ibaraki 305, Japan} \\
\r {42} {\eightit Tufts University, Medford, Massachusetts 02155} \\
\r {43} {\eightit Waseda University, Tokyo 169, Japan} \\
\r {44} {\eightit University of Wisconsin, Madison, Wisconsin 53706} \\
\r {45} {\eightit Yale University, New Haven, Connecticut 06520} \\
\vspace{0.3in} 

November 13, 1999\\
\vspace{0.1in} 

(submitted to \prl)\\
\end{center}

%\date{\today}
\maketitle

\clearpage
\begin{abstract}  
We have searched for direct production of scalar top quarks at the Collider
Detector at Fermilab in 88 pb$^{-1}$ of $p\overline{p}$ collisions at
$\sqrt{s}=$1.8 TeV.  We assume the scalar top quark decays into either a
bottom quark and a chargino or a bottom quark, a lepton, and a scalar
neutrino.  The event signature for both decay scenarios is a lepton, missing
transverse energy, and at least two $b$-quark jets.  For a chargino mass of
90 GeV/c$^2$ and scalar neutrino masses of at least 40 GeV/c$^2$, we find no
evidence for scalar top production and present upper limits on the production
cross section in both decay scenarios. 
\end{abstract}
\vspace{0.2in}
\vspace{0.2in}

\narrowtext

The Minimal Supersymmetric extension to the Standard Model (MSSM) \cite{susy}
assigns a scalar supersymmetric partner for every Standard Model fermion and
a fermionic superpartner for every Standard Model boson.  
The weak eigenstates of each scalar superpartner mix, forming mass eigenstates
\cite{baer}.  The splitting of the mass eigenvalues is proportional to the
mass of the Standard Model partner.  Therefore, the superpartners
of the top quark weak eigenstates, $\tilde{t}_L$ and $\tilde{t}_R$, may have
the largest mass splitting of all the scalar quarks (squarks).  The running
of the squark mass parameters is proportional to the Yukawa coupling of
the Standard Model partners, such that the diagonal elements of the
$\tilde{t}_L$,$\tilde{t}_R$ mass matrix should be smaller than those of the other squarks \cite{baer}.  Thus, the lighter
scalar top mass eigenstate, $\stop$, is the best candidate for the lightest
squark and is potentially lighter than the top quark.  
We report the results of a search for direct production of $\stop\stopb$ in  
88$\pm$4 pb$^{-1}$ of data collected during the 1994-95 Tevatron run using
the Collider Detector at Fermilab (CDF).

The CDF detector has been described elsewhere \cite{cdf}.
In this analysis, we used electrons
identified in the central 
electromagnetic calorimeter which covers the pseudorapidity region 
$|\eta|<$1.1.
We used muons identified by tracks in
drift chambers in two detector subcomponents outside the calorimeters.  The
first muon subsystem is located behind five absorption lengths of material
and covers the region $|\eta |<$0.6.  The second is located behind an
additional three absorption lengths of material and has the same $\eta$
coverage as the first.  

Scalar tops are strongly produced in the Tevatron via $q \overline{q}$
annihilation and gluon-gluon fusion.  We searched for $\stop\stopb$ production within the framework of the MSSM
for the case where $m_\stop<m_t$.  We assumed R-parity \cite{baer} 
is conserved and restricted ourselves to two separate \stop decay modes~\cite{modes}.  
In the first, the decay\footnote{Unless otherwise noted, decay channels
  imply their charge conjugates.} $\tilde{t}_1 \rightarrow b
\tilde{\chi}_1^+$, where $\tilde{\chi}_1^+$ is the lightest chargino,
proceeds with a branching ratio of 100\%.  We required one of the charginos,
which decay via a virtual $W$, to decay as $\tilde{\chi}_1^+ \rightarrow
e^+ \nu ~ \tilde{\chi}_1^0$ or $\mu^+ \nu ~ \tilde{\chi}_1^0$, where
$\tilde{\chi}_1^0$ is the lightest neutralino, with an assumed branching
ratio of 11\% for each lepton type.  
For models where $\tilde{t}_1
\rightarrow b \tilde{\chi}_1^+$ is not kinematically allowed, we considered a
second decay scenario in which $\stop \rightarrow b l^+\tilde{\nu}$, where
$\tilde{\nu}$ is a scalar neutrino and each $l=e,~\mu,~\tau$ has a branching
ratio of 33.3\%. 
In these two scenarios, either the $\tilde{\chi}_1^0$ or the $\tilde{\nu}$ is
the lightest supersymmetric particle (LSP) and does not decay.
A third possible decay scenario in which the $\stop \rightarrow c
\tilde{\chi}_1^0$ branching ratio is 100\% is the subject of separate CDF
searches \cite{holck}. 

In both decay scenarios considered here, the \stop signature is at least
one isolated lepton, missing transverse energy (\MET) from the neutral LSP's
and at least two jets from the $b$ quarks.  This signature is very similar to
that of the top quark, with kinematic differences due to the smaller \stop
mass in our search region, the presence of two massive neutralinos in the
final state, and the absence of a real W in the final state.  We therefore
expect events with lower lepton $p_T$, lower jet $E_T$ and multiplicity, and
without a peak in the lepton-\MET transverse mass.  To remain efficient for
the smaller \stop mass, we used data collected with the low-$p_T$ electron
and muon triggers described in Ref.\ \cite{blife}.  These trigger thresholds
were $E_T\ge$ 8 GeV for electrons and $p_T\ge$ 8 GeV/c for muons.

The data for this analysis were obtained by requiring (i) an electron with
\mbox{$E_T\ge$ 10 GeV} or muon with \mbox{$p_T\ge$ 10 GeV/c} originating
from the primary vertex and passing lepton identification cuts, (ii)
$\MET\ge$ 25 GeV, and (iii) at least two jets with cone sizes of $R \equiv
\sqrt{(\Delta\eta)^2 + (\Delta\phi)^2}=0.7$, one with $E_T \ge$ 12 GeV and
the second with $E_T\ge$ 8 GeV.  The lepton identification cuts were
identical to those used in previous CDF analyses \cite{top,trilep}.  For
electron identification, 
the electron was required 
to have lateral and longitudinal shower profiles
consistent with those of an electron, have less than 5\% of its energy
deposited in the hadronic calorimeter, and be well-matched to a track from
the CTC.  A muon was required to have tracks in the inner and outer central
muon chambers which were well-matched to a track from the CTC.  We further
required the leptons to pass an isolation cut in which the calorimeter $E_T$
in a cone of $R=0.4$ around the lepton was less than 2 GeV (excluding the
lepton tower).  No explicit tau identification was conducted, but leptons
from tau decays were accepted. 

We used the SVX$^\prime$ detector to identify secondary vertices from $b$ quark decays
and selected events with at least one secondary vertex.  The tagging
algorithm is described in Ref.\ \cite{top} with improvements given in Ref.\
\cite{top2} and efficiency measured in Ref.\ \cite{top_prod}.  We reduced the Drell-Yan background in our
sample by removing events with two isolated, opposite-sign leptons.
This background was further reduced by removing events with
an isolated lepton that reconstructed an invariant mass $\ge$ 50 \mgev\
with any additional, isolated
CTC track.  Finally, we
reduced the background from $b \overline{b}$ events and events with hadrons
misidentified as leptons (fake leptons) by requiring that the $\Delta\phi$
between the \MET and the nearer of the two highest-$E_T$ jets be $\ge$ 0.5
rad.  This reduces fake \MET due to jet energy
mismeasurement.  The number of events remaining in our sample after all cuts
is 81. 

Signal and background selection cut efficiencies were estimated using a
variety of Monte Carlo generators followed by a CDF detector simulation.
Signal event samples were created 
using \isajet\ version 7.20 \cite{isajet}.  The supersymmetric particle masses
used in signal simulation were:
$m_{\tilde{\chi}_1^\pm}=$ 90 \mgev, $m_{\tilde{\chi}_1^0}=$ 40 \mgev, 
and $m_{\tilde{\nu}}\ge$ 40 \mgev,
which are consistent with current lower limits~\cite{limits}.
The signal selection efficiency increases with $m_\stop$ but decreases with 
$m_{\tilde{\nu}}$ and $m_{\tilde{\chi}_1^\pm}$, reaching a plateau as event 
energies advance from cut thresholds.  Some specific efficiencies are 5.4\% 
for $\stop \rightarrow b l^+ \tilde{\nu}$ ($m_\stop=130$ \mgev, 
$m_{\tilde{\nu}}$ = 40 \mgev) and 0.7\% for $\stop \rightarrow b 
\tilde{\chi}_1^+$ 
($m_\stop=120$ \mgev, $m_{\tilde{\chi}_1^\pm} = 90$ \mgev, and 
$m_{\tilde{\chi}_1^0} = 40$ \mgev).  These selection efficiencies include 
branching ratios of forced decays.

The significant sources of uncertainty for signal selection efficiency are
(i) the $b$-jet tagging efficiency,
(ii) the trigger efficiencies,
(iii) the luminosity, and
(iv) initial- and final-state radiation.
The effects of some of these sources
vary with $m_\stop$, but none contribute more than 10\% to the
overall uncertainty, which is less than 16\%
for all $m_\stop$ considered.

Standard Model backgrounds come from any process that can produce two or more
jets, either real or fake leptons, and real or fake $\MET$.  This includes
heavy flavor quark production, vector boson production with two or more
accompanying jets, and inclusive jet production with real or fake leptons.
The number of events from the first two processes that we expected in our
data sample were predicted using measured or calculated cross sections and
selection efficiencies determined from Monte Carlo.  Top-pair and single-top
production were simulated using \herwig\ version 5.6 \cite{herwig}.  For 
$m_t$ = 175 \mgev\, $\sigma_{t \bar{t}}$ is 5.1 $\pm$ 1.6 ~{\rm pb}
\cite{top_prod} and $\sigma_{t \bar{b}}$ for $W$-gluon fusion from 
a next-to-leading-order (NLO) calculation is 1.70
$\pm$ 0.15 ~{\rm pb} \cite{stelzer}.  Vector boson samples 
were generated using \vecbos ~version 3.03 \cite{vecbos} and normalized
according to CDF measurement~\cite{properties_of_Ws}.  Drell-Yan, $b
\bar{b}$, and $c \bar{c}$ samples were generated with \isajet\ version 7.06
and normalized to independent CDF data samples.  

To determine the number of events with fake leptons in our sample, we used
a data sample passing all our selection cuts with the exceptions of a
non-overlapping \MET requirement ($15\le \MET \le 20$ GeV) and no requirement
on $\Delta\phi$($\MET,$ nearer jet).  The number of fake lepton events was
normalized to this data sample, which contained negligible signal, after
other backgrounds were subtracted.  The number of fake lepton events was then
extrapolated to the signal region using cut efficiencies determined from an
independent fake-lepton event sample.   

The complete list of backgrounds and the number of expected events remaining
after all cuts is given in Table \ref{after}.  The significant backgrounds
are $t\overline{t}$, $b \overline{b}$, $W^\pm (\rightarrow l^\pm\nu) +\ge$2 jets,
and fake lepton events.  The number of data events agrees well with the
expected background. 

To determine the number of potential signal events
in this final data sample, we performed extended, unbinned likelihood fits for
each \stop mass considered for both decay scenarios.  
The likelihood fits compared the shapes of distributions of the 
signal and background
and included Gaussian terms tying the fit background levels to their
predicted levels.
The fit parameters were the numbers of signal events, 
$t\overline{t}$ events, $b \overline{b}$ plus
fake lepton events, 
and vector boson events
(represented in the fit by the $W^\pm +\ge$2 jets distributions).
We used the Kolmogorov statistic applied to the simulated distributions of
signal and 
combined 
backgrounds to determine the most sensitive kinematic
distributions to use in the fit.  The kinematic distributions
evaluated include lepton $p_T$, 
$H_T$ (the scalar sum of 
lepton $E_T$, \MET, and jet $E_T$ for all
jets with $E_T\ge$ 8 GeV), jet multiplicity, and $\Delta \phi$(jet1,jet2),
where jets are ordered in $E_T$.

\begin{figure}
\epsfig{file=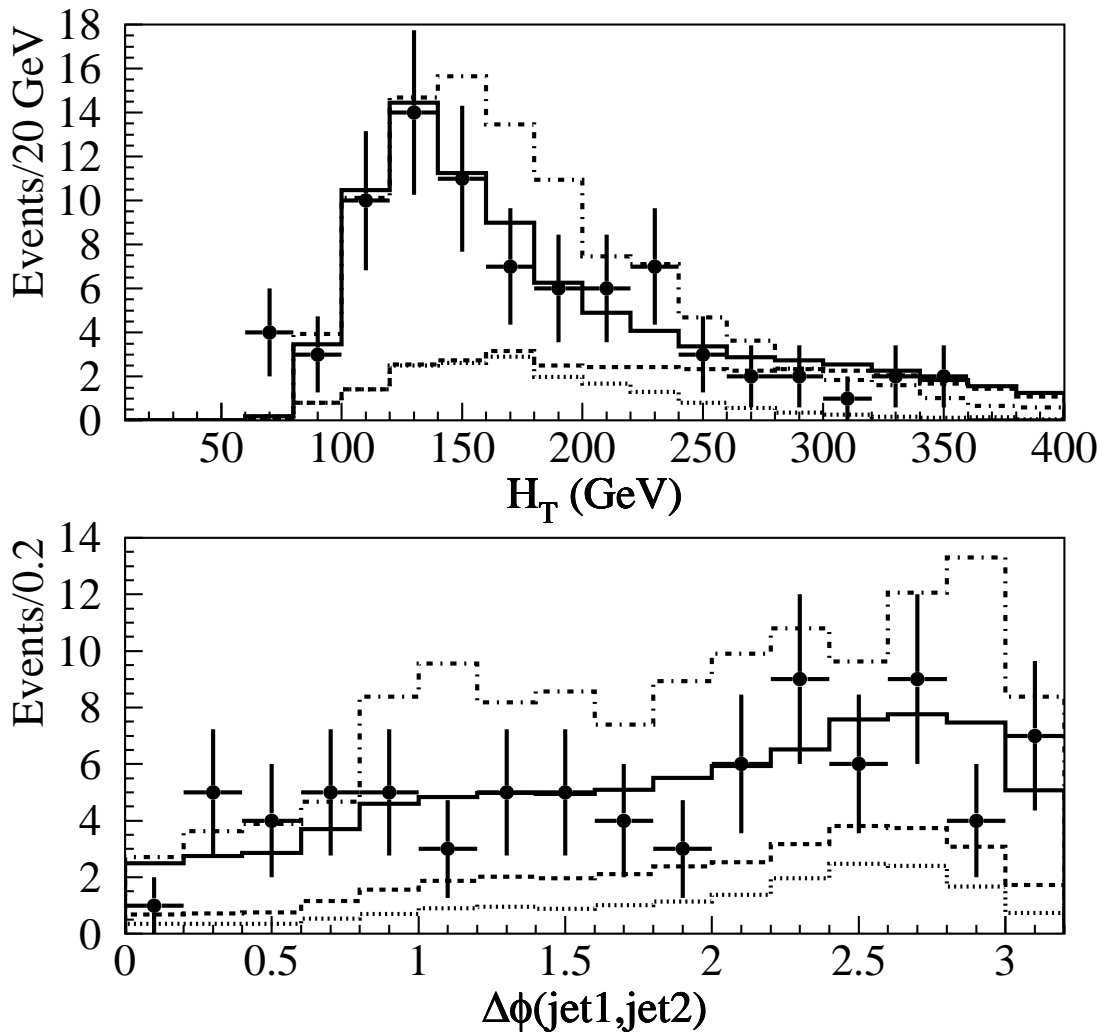,height=16cm}
{\caption
{\label{fig:afterfit} Results of the two-dimensional fit to $H_T$ and $\Delta
\phi$(jet1,jet2) when the $\tilde{t}_1 \rightarrow b
\tilde{\chi}_1^+$ branching ratio is 100\%, $m_\stop$ = 115 \mgev,
$m_{\tilde{\chi}_1^\pm}$ = 90 \mgev, and $m_{\tilde{\chi}_1^0}$ = 40 \mgev.  
The quantities $H_T$ and $\Delta
\phi$(jet1,jet2) are defined in the text.  The points represent the data.
Cumulative contributions from $b \overline{b}$ and fake lepton events, $t
\overline{t}$, and $W^\pm\rightarrow l^\pm \nu$ + jets are represented by
dotted, dashed and solid lines respectively. 
There is no significant contribution from signal.  To
illustrate the shape difference, a signal distribution with arbitrary normalization has
been overlaid with a dot-dash line. }}
\end{figure}

\begin{figure}
\epsfig{file=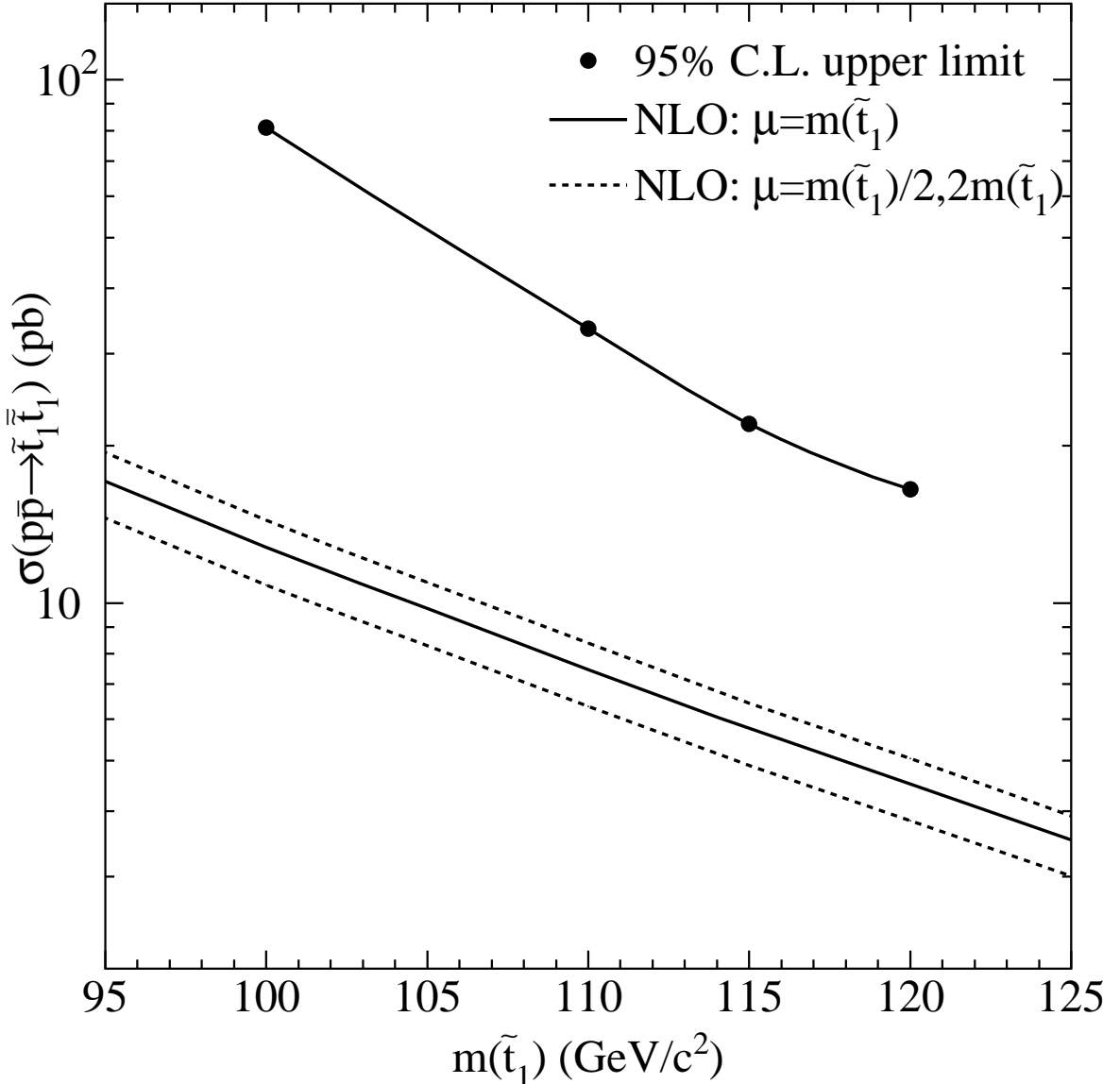,height=17cm}
{\caption[Cross section limit for \stop in $\tilde{\chi}_1^+$ decay scenario.]
{\label{fig:limit} The points represent the CDF 95\% C.L. cross section limit 
as a function of \stop mass when the $\tilde{t}_1 \rightarrow b
\tilde{\chi}_1^+$ branching ratio is 100\%, 
$m_{\tilde{\chi}_1^\pm}=$ 90 \mgev, and $m_{\tilde{\chi}_1^0}=$ 40
\mgev. The line without markers represents the
NLO prediction for $\sigma_{\stop\stopb}$ using the renormalization
scale $\mu=m_\stop$.
The dashed lines represent the
NLO cross section for $\mu=m_\stop/2$ and $\mu=2 m_\stop$. }}
\end{figure}

For the $\tilde{t}_1 \rightarrow b \tilde{\chi}_1^+$ decay, 
sensitivity to signal was greatest for a two-dimensional 
fit to the combined probability distributions for $H_T$ and
$\Delta \phi$(jet1,jet2).  
Fit results at all masses were consistent with zero signal events.
The fit result for $\tilde{t}_1 \rightarrow b
\tilde{\chi}_1^+$ with $m_\stop=115$ \mgev, $m_{\tilde{\chi}_1^\pm}$ = 90
\mgev, and $m_{\tilde{\chi}_1^0}$ = 40 \mgev\ is shown in Fig.\ \ref{fig:afterfit}.
The 95\% C.L. limits on $\sigma_{\stop\stopb}$ for this decay
are shown in Fig.\ \ref{fig:limit} as a function of $m_\stop$ \cite{feldman}.  The
NLO prediction for $\sigma_{\stop\stopb}$ using the renormalization
scale $\mu=
m_\stop$ and parton distribution function CTEQ3M is shown in Fig.\
\ref{fig:limit} for comparison \cite{beenakker}.

\begin{figure}
\epsfig{file=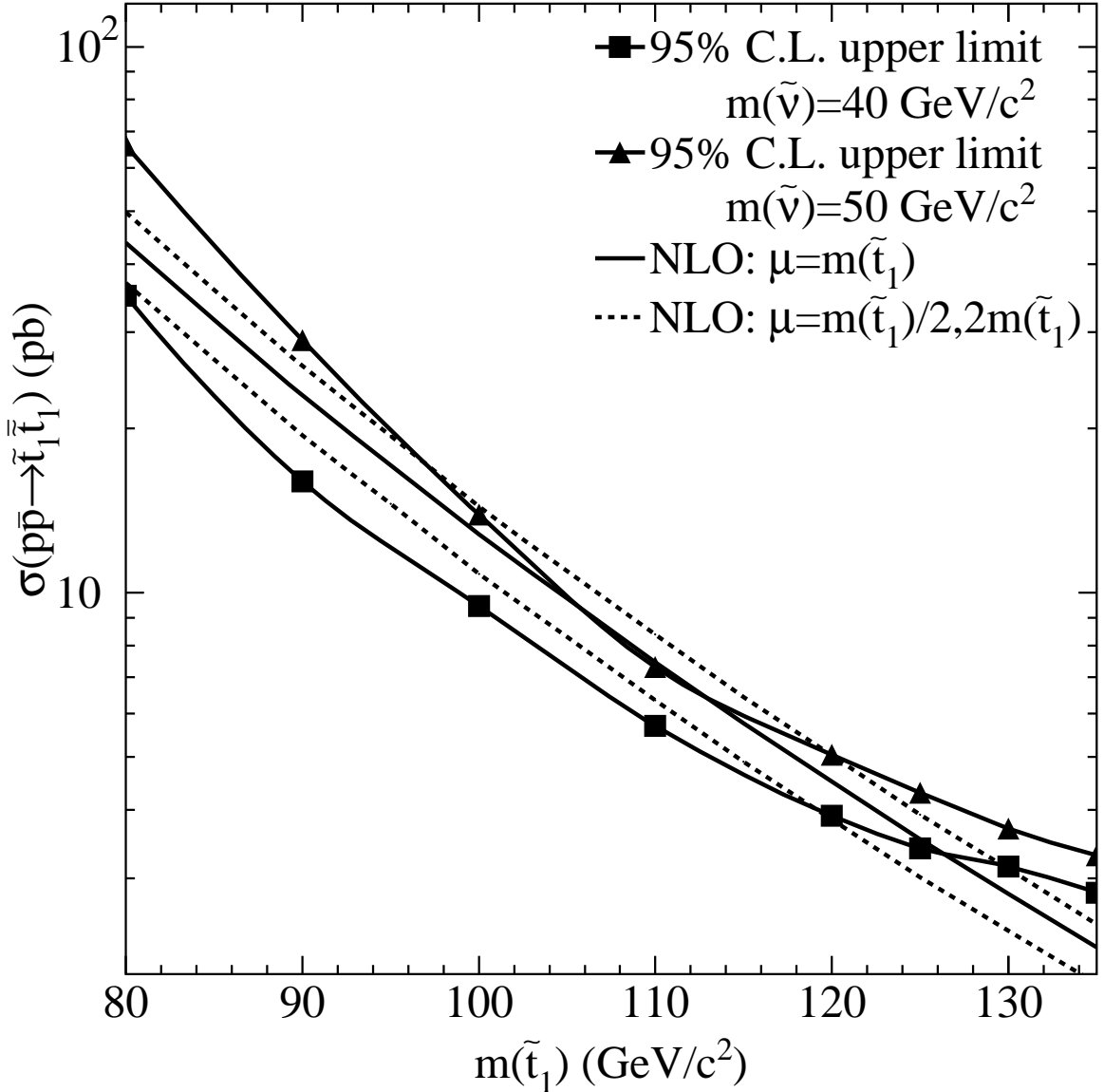,height=17cm}
{\caption
{\label{fig:limit1} CDF 95\% C.L. cross section limit as a function of
\stop mass when the $\stop \rightarrow b l^+\tilde{\nu}$ branching ratio is
100\% and $m_{\tilde{\nu}}=$ 40 \mgev\ (squares) or 50 \mgev\ (triangles).
The line without markers represents the
NLO prediction for $\sigma_{\stop\stopb}$ using the renormalization
scale $\mu=m_\stop$.
The dashed lines represent the
NLO cross section for $\mu=m_\stop/2$ and $\mu=2 m_\stop$. }}
\end{figure}

\begin{figure}
\epsfig{file=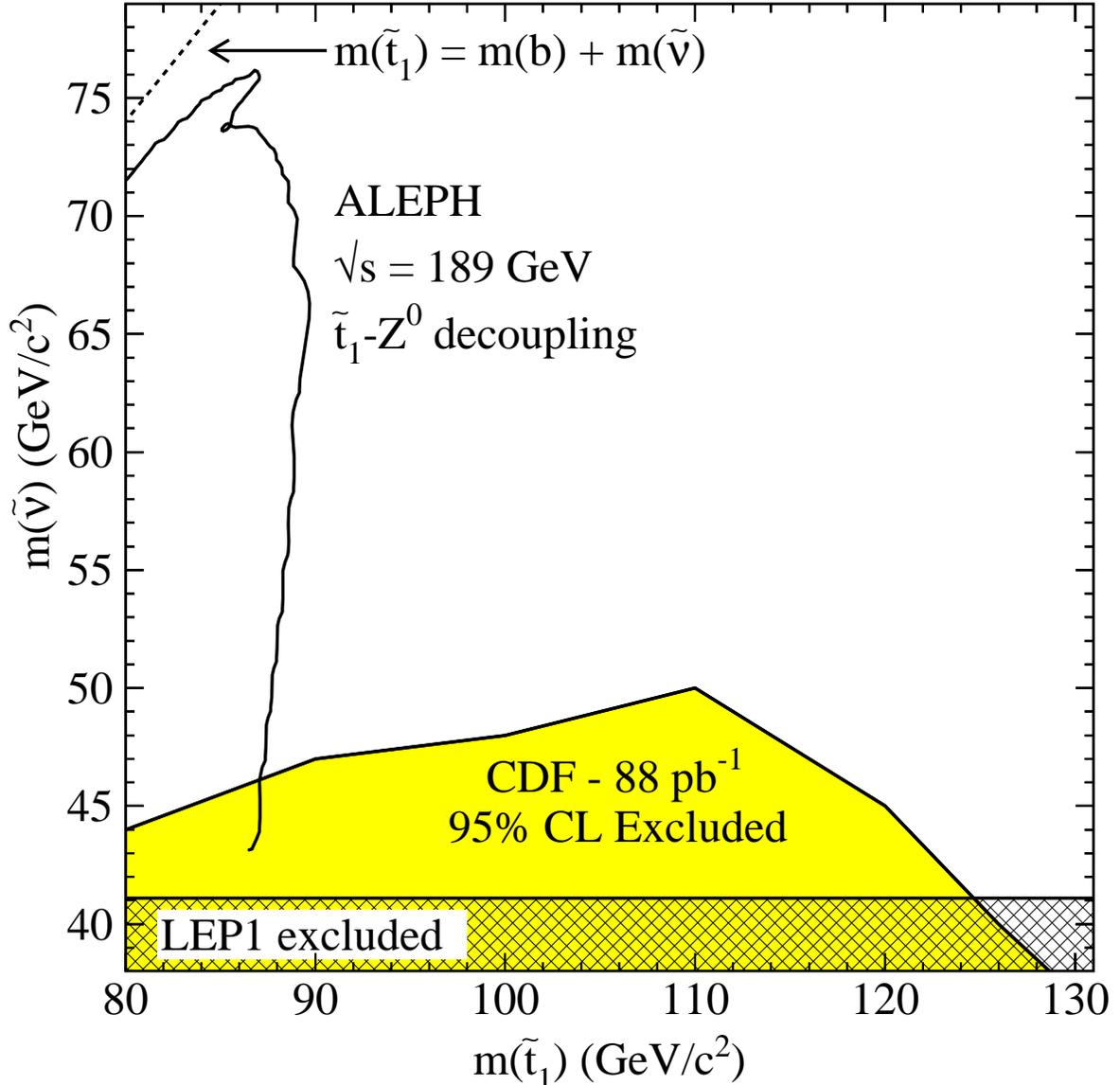,height=17cm}
\caption{95\% C.L. excluded region in the plane of $m_\stop$
versus $m_{\tilde{\nu}}$ when the $\stop \rightarrow b e^+
\tilde{\nu}$, $\stop \rightarrow b \mu^+ \tilde{\nu}$, and $\stop 
\rightarrow b \tau^+ \tilde{\nu}$ branching ratios are 
33.3\%.  We define the exclusion region as that region 
of supersymmetric parameter space for which the 95\% C.L. limit on
$\sigma_{\stop\stopb}$ is less than the NLO prediction ($\mu=
m_\stop$).  The LEP1 $m_{\tilde{\nu}}$ limit and ALEPH excluded region in the $m_\stop$
versus $m_{\tilde{\nu}}$ plane are also shown \protect\cite{aleph}.  The
ALEPH excluded region corresponds to the case in which the \stop decouples
from the $Z^0$.} 
\label{fig:exclu}
\end{figure}

For the $\stop \rightarrow b l^+\tilde{\nu}$ decay scenario,
sensitivity to signal
was greatest for a fit to 
the $H_T$ distribution.  Again, all fit results were consistent 
with zero signal events.
The 95\% C.L. limits on
$\sigma_{\stop\stopb}$ for the $\stop \rightarrow b l^+
\tilde{\nu}$ decay are shown in Fig.\ \ref{fig:limit1} for
$m_{\tilde{\nu}}$ = 40 and 50 \mgev. 
We consider the regions of supersymmetric parameter space
for which the 95\% C.L. limit on
$\sigma_{\stop\stopb}$ is less than the NLO prediction ($\mu=
m_\stop$) to be
excluded.  The resulting excluded
region in the plane of $m_\stop$ versus $m_{\tilde{\nu}}$ for is shown in Fig.\ \ref{fig:exclu}.

To conclude, we have searched for direct $\stop\stopb$ production in 
88$\pm$4 pb$^{-1}$ 
of data collected using the CDF detector during the
1994-95 Tevatron run.
We found no evidence for $\stop\stopb$ production for either 
$\tilde{t}_1 \rightarrow b \tilde{\chi}_1^+$ or $\stop
\rightarrow b l^+\tilde{\nu}$ and present upper limits on
$\sigma_{\stop\stopb}$ as a function of $m_\stop$.

We thank the Fermilab staff and the technical staffs of the participating
institutions for their vital contributions.  This work was supported by the
U.S. Department of Energy and National Science Foundation; the Italian
Istituto Nazionale di Fisica Nucleare; the Ministry of Education, Science
and Culture of Japan; the National Sciences and Engineering Research
Council of Canada; the National Science Council of the Republic of China;
the A. P. Sloan Foundation; and the Swiss National Science Foundation.

\begin{table}
\caption{
Number of data events and expected background events after all selection cuts.
The dominant sources of uncertainty on the 
numbers of expected events are integrated luminosity, cross section, trigger 
efficiency, and b-jet tagging.
(Fake leptons are hadron tracks which have been misidentified as leptons.)}
\label{after}
\begin{center}
\begin{tabular}{lc}
process&number of events\\
&expected after all cuts\\ \hline 
$W^\pm (\rightarrow e^\pm\nu $ or $ \mu^\pm\nu)+\ge$2 jets&$44.5 \pm 7.3~\,$ \\
$t \overline{t}$                             &$17.8 \pm 4.5~\,$ \\
$b \overline{b}$                             &$~5.8 \pm 0.8~$ \\   
$W^\pm (\rightarrow \tau^\pm\nu )+\ge$2 jets   &$~2.6 \pm 0.4~$ \\
$t \overline{b}$ (from $W-g$ fusion)         &$~1.6 \pm 0.2~$ \\
$Z (\rightarrow e^+e^- $ or $ \mu^+\mu^- )+\ge 2$ jets&$~1.4 \pm 0.2~$ \\
$Z (\rightarrow \tau^+\tau^-)+\ge 1$ jet      &$~0.4 \pm 0.1~$ \\
$\gamma \rightarrow l^+l^-$                  &$~0.4 \pm 0.1~$ \\
$c \overline{c}$                             &$0.06 \pm 0.02$\\   
fake lepton events                           &$12.7 \pm 1.6~\,$ \\ \hline
background total                             &$87.3 \pm 8.8~\,$ \\ \hline \hline
data                                         &      81     \\
\end{tabular}
\end{center}
\end{table}

\end{document}